\documentclass[a4paper]{article}

\usepackage{icrc2013}
\usepackage{url}
\usepackage{unites2e}
\usepackage[english]{babel}

\title{Ground-based Gamma Ray Astronomy}

\shorttitle{Ground-based Gamma Rays}

\authors{
Jamie Holder$^{1}$
}

\afiliations{
$^1$ Department of Physics and Astronomy and the Bartol Research
Institute, University of Delaware, Newark, Delaware, USA. \\
}

\email{jholder@physics.udel.edu}

\abstract{This paper is the write-up of a rapporteur talk given by the
  author at the 33rd International Cosmic Ray Conference in Rio de
  Janeiro, Brazil, in 2013. It attempts to summarize results and
  developments in ground-based gamma-ray observations and
  instrumentation from among the $\sim300$ submissions to the
  gamma-ray sessions of the meeting. Satellite observations and
  theoretical developments were covered by a companion rapporteur
  \cite{Stawarz_review}. Any review of this nature is unavoidably
  subjective, and incomplete. Nevertheless, the article should provide
  a useful status report for those seeking an overview of this
  exciting and fast-moving field.}


\begin{document}
\maketitle

\section*{Introduction}

Over the past decade, gamma-ray astronomy has risen to prominence as
probably the most productive sub-field of astroparticle physics. This
is due to the results of the current generation of
ground-based and space-based instruments, which have provided the
first sensitive view of the gamma-ray sky. The importance of the field
was recognized at the 33rd International Cosmic Ray Conference (ICRC)
in Rio de Janeiro, Brazil, by a modification to the organization of
the program, which now includes dedicated gamma-ray sessions
sub-divided into experimental (GA-EX), intrumentation (GA-IN) and
theoretical (GA-TH) branches. Around 100 talks and more than 200
poster presentations were given in the gamma-ray sessions at the
conference. This paper attempts to summarize some of the highlights
from these, with a focus on ground-based gamma-ray observations and
instrumentation. A companion paper by L. Stawarz will cover space and
ballloon-based gamma-ray observations, and gamma-ray astronomy theory
\cite{Stawarz_review}. A broad review of the field was also presented
at the conference by J. Hinton \cite{Hinton_review} Slides from the
associated rapporteur presentation are also available for download
from the conference INDICO site
\footnote{\url{http://143.107.180.38/indico/contributionDisplay.py?contribId=1310&sessionId=8&confId=0}}.

At the time of the previous ICRC, held in Beijing in 2011, the TeV
catalogue contained 132 sources. The 2011 conference saw a number of
ground-based highlights \cite{funk_rapp}: the detection of the Crab
pulsar above $100\U{GeV}$ by VERITAS and MAGIC challenged models of
high energy emission from pulsars \cite{crab_beijing_VTS,
  crab_beijing_MAGIC}. H.E.S.S. reported the likely detection of TeV
gamma-ray emission from a globular cluster \cite{terzan5_beijing}, and
the first pulsar wind nebula outside of our own Galaxy
\cite{LMC_PWN_beijing} . The addition of Fermi-LAT data to
ground-based measurements of the SNR RX~J1713.7-39466 showed strong
support for a leptonic orgin \cite{RXJ1713_beijing}, while the
broad-band spectrum of Tycho's SNR favoured a hadronic scenario
(although with much more limited statistics)
\cite{tycho_beijing}. Blazars dominated the extragalactic results,
which were marked by a dramatically increased rate of new detections
following of the release of the first Fermi-LAT catalogue, and the
development of strictly contemporaneous, broad-band datasets, aided by
extensive cross-collaboration coordination and cooperation
\cite{funk_rapp}. The major instrumentation development was the
commissioning of the second $17\U{metre}$ MAGIC telescope, from which
first results were presented \cite{MAGICII_beijing}.

\begin{figure*}[!]
  \centering
  \includegraphics[width=\textwidth]{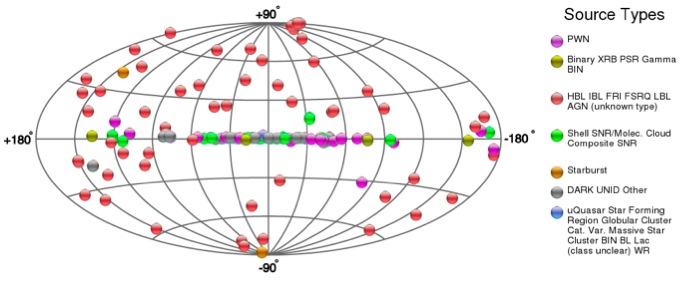}
  \caption{The TeV source catalog in Galactic coordinates as of summer
    2013, courtesy of TeVCat.}
  \label{TeVCat}
\end{figure*}
 
In the two years since Beijing, the field has further progressed, and
new discoveries continue to be made. Figure~\ref{TeVCat} shows the TeV
sky as of summer 2013, containing now 145
sources\footnote{\url{http://tevcat.uchicago.edu/}}. I will attempt to
summarize some of the more exciting new scientific results in these
proceedings, but a clear theme of the 2013 meeting was the
commissioning of new instruments, the successful completion of major
upgrades to many of the existing facilities, coupled with a dramatic
ramp-up in development work towards the next generation of
instruments. This bodes extremely well for the near, medium and
long-term future of the field. I begin
by summarizing the current instrumental status.

\section*{Imaging Atmospheric Cherenkov Telescopes}

Three major Cherenkov telescope arrays are operating around the world
at present: VERITAS, H.E.S.S. and MAGIC, together with a number of
smaller instruments. The three large facilities are all running
smoothly, and have recently undergone major upgrades. The stereoscopic
imaging technique they employ provides the highest sensitivity for
gamma-ray astronomy from the ground, together with the best angular
and energy resolution. Given the limited fields-of-view
($\leq5^{\circ}$), observations must be targeted, but the three arrays
are distributed around the globe in such a way as to provide
reasonable coverage of the entire sky. All provided detailed status
updates in highlight talks at the ICRC \cite{VERITAS_status,
  HESS_status, MAGIC_status} and I briefly summarize their status
below, in an arbitrary reverse alphabetical order.

\subsubsection*{VERITAS}  
The VERITAS array of four $12\U{m}$ diameter telescopes, located in
Arizona, USA, has been in operation since the summer of 2007. Summer
2012 saw the completion of a program of major upgrades to the
facility. These included the relocation of the initial prototype
telescope to a more favorable location in the array (in 2009), and the
installation of faster telescope-level trigger systems and a
fiber-optic network (in 2011/12). A major development over the past
two years was the replacement of all of the telescope photodetectors
with more sensitive, ``super-bialkali'' devices. This was completed in
summer 2012, resulting in a 50\% increase in the Cherenklov light
yield, a 30\% reduction in energy threshold and improved sensitivity,
particulary for soft spectrum sources such as distant blazars
\cite{Kieda, VERITAS_status}.  All of the VERITAS contributions to the
conference are referenced on the arXiv in \cite{VERITAS}.

\subsubsection*{MAGIC-II} 
MAGIC has been observing from La Palma since 2004, initially with a
single $17\U{m}$ diameter telescope, and, since 2009, with a stereo
pair. Further upgrades have recently been completed
\cite{MAGICII_upgrade}, including the installation of an entirely new
camera and trigger system for the original telescope, which now
matches that of the second telescope and provides a homogeneous system
\cite{MAGICII_cam}. The data acquisition system for both telescopes
has also been replaced and standardized with a system based on the
DRS4 chip \cite{MAGICII_readout}. The system is performing well,
providing an integral sensitivity of $0.71\pm0.02\%$ of the Crab
Nebula flux for a 50 hour observation\cite{MAGIC_performance}.  Future
plans include the installation of an analog sum trigger system in fall
2013 for both telescopes, which will operate in parallel with the
existing system and further lower the energy threshold
\cite{MAGIC_sumtrigger}.

\subsubsection*{H.E.S.S. and H.E.S.S. II}

\begin{figure*}[!t]
  \centering
  \includegraphics[width=\textwidth]{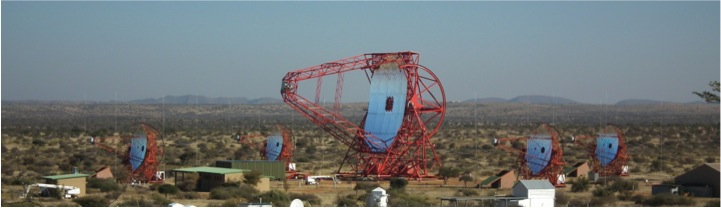}
  \caption{The H.E.S.S. II array in Namibia, showing the four original
    $12\U{m}$ diameter telescopes and the newly commissioned $28\U{m}$
    diameter telescope in the centre \cite{HESS_status}}
  \label{HESS2}
\end{figure*}
 
The H.E.S.S. array of four, $12\U{m}$ telescopes in Namibia was the
first of the current generation of instruments to begin operations,
and celebrated its 10th anniversary in 2012 \cite{HESS_status}. The
array has taken almost $10,000\U{hours}$ of data, including
$2,800\U{hours}$ on a scan of the Galactic Plane \cite{HESS_GPS}. The
plane scan is now complete, and future observations will target
individual objects. These studies will be greatly enhanced by the
addition of H.E.S.S. II, a $614\UU{m}{2}$ telescope located at the
centre of the array and equipped with a 2048-pixel photomultiplier
tube camera (Figure~\ref{HESS2}). H.E.S.S. II is a technical
\textit{tour de force} - the largest Cherenkov telescope ever
constructed - and the mechanical performance seems good
\cite{HESSII_focus, HESSII_drive}.  The collaboration presented first
light at the ICRC, with a detection of the Crab Nebula
\cite{HESS_status}.

\subsubsection*{FACT}

\begin{figure}[!]
  \centering
  \includegraphics[width=0.48\textwidth]{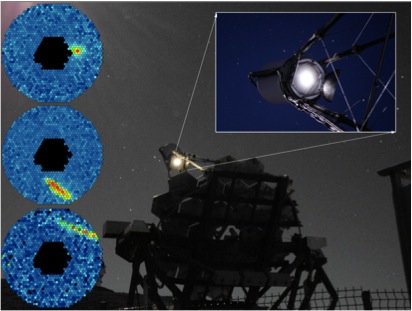}
  \caption{Observations of cosmic ray events with the full moon in the
  centre of the FACT field of view \cite{FACT_status}.}
  \label{FACTMoon}
\end{figure}

A relative newcomer to the field is FACT (First G-APD Cherenkov
Telescope), located on La Palma and operating since 2011
\cite{FACT_status}. As a single telescope with modest aperture
($9.5\UU{m}{2}$), FACT is scientifically limited to (useful)
monitoring of the brightest TeV blazars \cite{FACT_AGN}. However, it
also serves as an important testbed for some of the technologies which
will be employed in the next generation of Cherenkov instruments. The
1440 pixel silicon photodetector camera is the clearest example of
this, and results confirming its stability and operation under bright
moonlight conditions were presented \cite{FACT_moonlight,
  FACT_stability} (Figure~\ref{FACTMoon}). The telescope is also
routinely operated remotely, with no need for on-site observers, with
an ultimate goal of completely automatic, robotic operation
\cite{FACT_robot}.

\section*{Non-imaging Cherenkov \& Particle Detectors}

Gamma-ray astronomy from the ground can also be performed by sampling
the Cherenkov light pool with non-imaging detectors, or by measuring
shower particles with detectors placed at high altitude. While the
sensitivity, angular and energy resolution of these techniques are
limited, in comparisoin with the imaging telescopes, this is offset by
the extremely wide field of view (typically $\sim2\U{sr}$) and, in the
case of particle detectors, close to 100\% duty cycle.

\subsubsection*{ARGO-YBJ} 
The ARGO-YBJ experiment was a unique air shower detector consisting of
a $74\U{m}\times78\U{m}$ carpet of Resistive Plate Chambers (RPCs)
surrounded by a partially instrumented area out to
$100\U{m}\times110\U{m}$, located at Yangbajing (Tibet, China) at an
altitude of $4300\U{m}$ above sea level. Data-taking concluded after 5
years of observations in February 2013, with an integral sensitivity
for the total dataset of $\sim25\%$ of the Crab Nebula flux
\cite{ARGO_status}. Summary results were presented at the ICRC - one
outstanding question remains the non-detection of the Milagro source
MGRO~J2019+37 by ARGO-YBJ, despite confirmation of this source with
VERITAS, and the detection of other Milagro sources with ARGO-YBJ
\cite{ARGO_extended}.

\subsubsection*{Tibet-III}
The Tibet air shower array also operates at Yangbajing, and consists
of a high density array of over 500 scintillation counters with
$7.5\U{m}$ spacing, together with wider spaced outriggers. The array
has been operating in various configurations since 1990, and provides
an angular resolution of $\sim0.9^{\circ}$ at $10\U{TeV}$
\cite{Tibet_status}. Results of a point source search for emission
were presented, showing significant emission only from the Crab and
Markarian 421 \cite{Tibet_point}, together with some evidence for
extended emission from Milagro sources \cite{Tibet_MGRO}. Cosmic ray
discrimination currently relies upon reconstructing the arrival
direction of the particle shower, but the addition of large
underground muon detectors is now in progress, and will allow
discrimination based upon the shower muon content
\cite{Tibet_MD}. Data-taking with 5 of the planned 12 muon detectors
will begin this year.

\subsubsection*{HAWC}

\begin{figure}[!]
  \centering
  \includegraphics[width=0.48\textwidth]{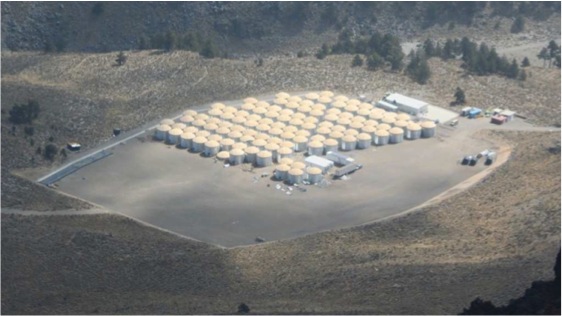}
  \caption{The High Altitude Water Cherenkov (HAWC) observatory on the
    Pico de Orizaba , with 100 tanks installed (one-third of the completed detector)}
  \label{HAWC}
\end{figure}

HAWC, the High Altitude Water Cherenkov observatory, is the successor
to Milagro, currently under construction at $4,100\U{m}$ above sea
level on the Sierra Negra volcano in Mexico. The final array will
consist of 300 water Cherenkov detectors, and provide order of
magnitude improvements in sensitivity, angular and energy resolution
over Milagro. 100 detectors had been deployed at the time of the ICRC,
as illustrated in Figure~\ref{HAWC} \cite{HAWC_status}. First results
from the array were shown, including measurements of the lunar cosmic
ray shadow using 30 tanks \cite{HAWC_moon} and the first evidence for
a gamma-ray source from two weeks of 77-tank observations of the Crab
Nebula \cite{HAWC_crab}. Deployment and commissioning are ongoing,
with science operations already underway, and completion of the full
array is expected by the end of 2014.

\subsubsection*{HiSCORE} 
Contributions from the HiSCORE collaboration were included in the
cosmic ray track at the ICRC \cite{HiSCORE} and so may have been
missed by members of the gamma-ray community. However, the primary
goal of this non-imaging Cherenkov detector is gamma-ray astronomy in
the $10\U{TeV}$ to several PeV range. A prototype array of three
wide-angle ($0.6\U{sr}$) optical stations has been deployed in the
Tunka Valley near Lake Baikal, and technical tests are underway. The
goal is to move towards a $1\UU{km}{2}$ engineering array with 60
stations over the next two years.

\section*{Results: Galactic Sources}
\subsubsection*{Galactic diffuse emission}
While diffuse emission from the Galactic plane dominates the sky at
$<100\U{GeV}$ energies, discrete sources are much more prominent at
the higher energies probed with ground-based instruments. Milagro have
reported the detection of a diffuse Galactic component, but a
substantial fraction of this may be simply explained by the sum of a
collection of unresolved sources below the Milagro detection threshold
\cite{Casanova08}. In truth, the distinction between very widely
extended sources and a diffuse background component is not strict. The
ARGO-YBJ collaboration presented confirmation of the Milagro measurement at
this ICRC \cite{ARGO_status}. H.E.S.S. have attempted to extract a
truly diffuse background component from their Galactic plane scan, in
an analysis which excludes any region containing a potential signal
from the background estimation \cite{HESS_diffuse}. The remaining flux
can be attributed to unresolved sources, and to cosmic ray
interactions with diffuse matter and photon fields. Gamma rays from
pion decay resulting from cosmic ray - matter interactions must
comprise at least 25\% of this diffuse flux. One conclusion to draw
from this study is that, while the analysis is complex, imaging
atmospheric Cherenkov telescopes are certainly capable of measuring
widely extended and even diffuse sources, despite their limited field
of view.

\subsubsection*{Galactic Centre}

\begin{figure}[!]
  \centering
  \includegraphics[width=0.48\textwidth]{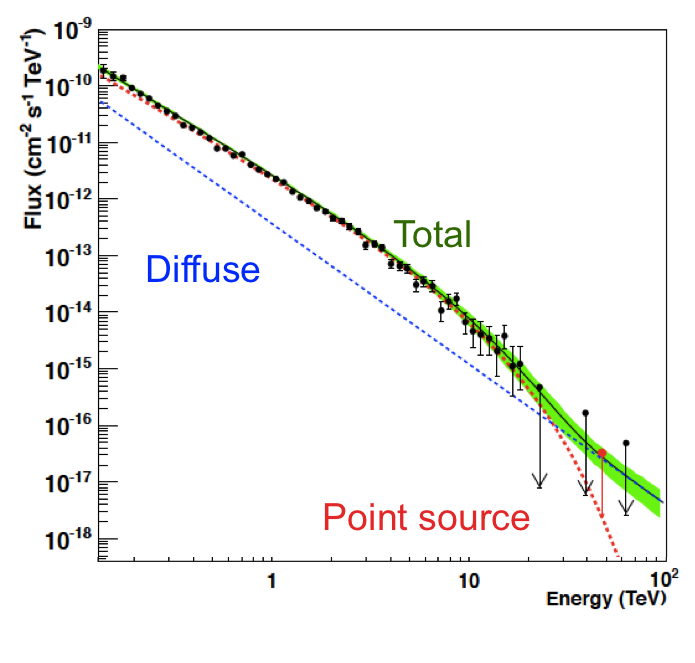}
  \caption{The results of a reanalysis of the Galactic Center TeV
    source, HESS J1745-290. Fitting the observed spectrum with diffuse and point
    source components lowers the point source spectral cut-off from 11
    to 7 TeV.}
  \label{HESS_GC_fig}
\end{figure}

The Galactic centre region hosts one known source of diffuse TeV
emission, the result of cosmic rays interacting with high density
molecular clouds along the Galactic centre ridge \cite{GC_ridge}. This
diffuse component also contributes to the emission observed from the
central point-like source in the direction of SgrA*. A new analysis of
the TeV point source was presented which attempts to remove this
component, estimated at $\sim20\%$ of the total emission, which
differs in spectral slope from the point source and displays no
spectral cut-off \cite{HESS_GC}. Accounting for this significantly
shifts the cut-off energy for the point source from $11\U{TeV}$ to
$7\U{TeV}$ (Figure~\ref{HESS_GC_fig}).

\subsubsection*{The Crab Pulsar and Nebula}

High energy studies of both the Crab Nebula and Pulsar have been
reinvigorated in recently years following the discovery of a very high
energy ($>100\U{GeV}$) component to the pulsed emission, and
variability in the ``steady'' flux at lower energies. The most extreme
example of this variability is in the Crab flares detected by AGILE
and Fermi-LAT, in which the high-energy synchrotron flux from the
nebula has been observed to increase by more than an order of
magnitude. No new details on the pulsar were presented by ground-based
instruments in the gamma-ray sessions, although some future exciting
results were hinted at in the MAGIC highlight talk: in particular the
fact that the pulsar peak widths (which are related to the geometry
and size of the emission region) are exceptionally narrow at the
highest energies \cite{MAGIC_status}.

\begin{figure}[!]
  \centering
  \includegraphics[width=0.48\textwidth, clip,trim = 0mm 0mm 1mm 0mm]{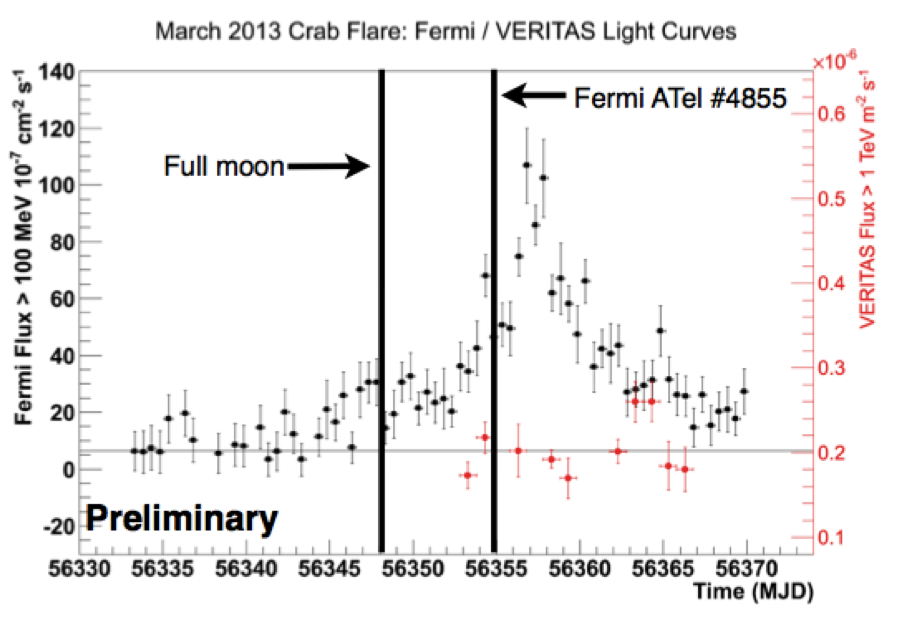}
  \caption{Fermi-LAT and VERITAS light curves for the March 2013 Crab
    Nebula flare. The baseline Crab Nebula synchrotron flux above $100\U{MeV}$ and average VHE flux above $1\U{TeV}$ are aligned, and are indicated by the solid black line. The vertical scales of both light curves have been adjusted such that the zero points and baseline fluxes are coincident.}
  \label{Crab_flare_fig}
\end{figure}

At this conference, ground-based observations of the flaring Crab were
presented by ARGO-YBJ, HAWC and VERITAS. ARGO-YBJ presented a
re-assessment of their archival observations of the Crab
\cite{ARGO_oldcrab}, which shows some evidence for a persistent
correlation between the ARGO and Fermi-LAT fluxes, and for a flux
enhancement of a factor of $2.4\pm0.8$ during bright flares. However,
they conclude that the low statistical significance of these results
does not allow to claim the detection of flux variability and requires
a confirmation by more sensitive instruments \cite{ARGO_crab}. In
March 2013 the Crab flared brightly again, reaching 20 times the
average steady synchrotron flux above $100\U{MeV}$
\cite{LAT_crab_2013}. HAWC was observing with 30 tanks operational, and
saw no flux enhancement - they estimate that a 5-fold flux increase
would have been detectable with $5\sigma$ significance
\cite{HAWC_crab}. VERITAS observations during the flare also show no
enhancement above $1\U{TeV}$ \cite{VERITAS_status}, and constrain
the variability at these energies to be less than 1\% of that seen by
the LAT (Figure~\ref{Crab_flare_fig}).

\subsubsection*{Other Pulsar Wind Nebulae}

\begin{figure}[!]
  \centering
  \includegraphics[width=0.48\textwidth]{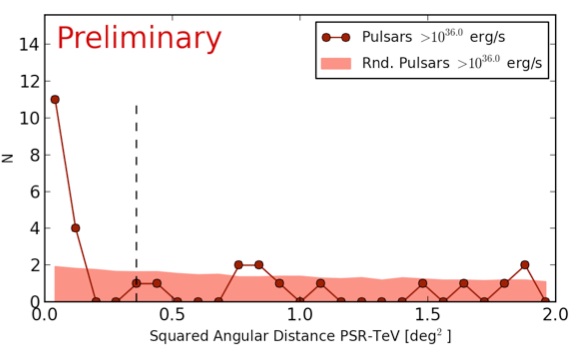}
  \caption{The number of high spindown energy pulsars, plotted as a function of
  their squared angular distance from a H.E.S.S. TeV source \cite{HESS_PWN}.}
  \label{HESS_PWN_fig}
\end{figure}

The Crab remains the only known source of pulsed gamma-rays above
$100\U{GeV}$ (indeed, the MAGIC Collaboration presented limits from an
unsuccessful $75\U{hour}$ search for pulsed emission from Geminga at
this conference \cite{MAGIC_Geminga}). Pulsar wind nebula, however,
are the most common class of Galactic TeV sources, which allows them
to be subjected to broad population studies. The
H.E.S.S. collaboration have begun such a study, using their Galactic
plane scan archive and comparing to pulsar characteristics in the ATNF
pulsar catalog \cite{HESS_PWN}. A consistent analysis has been applied
at each pulsar location in order to derive fluxes and upper
limits. This approach leads to a uniform selection and comparison of
candidates, which should allow meaningful study of PWN properties and
evolution. A simple first result of this study is the broad statement
that high \.{E} ($>10^{35} \U{erg}\UU{s}{-1}$) pulsars tend to have
TeV signals within $0.5^{\circ}$ of their location - there are no
spatial correlations beyond those expected from chance coincidence for
less energetic pulsars (Figure~\ref{HESS_PWN_fig}).

\begin{figure}[!]
  \centering
  \includegraphics[width=0.48\textwidth, clip,trim = 0mm 0mm 1mm 0mm]{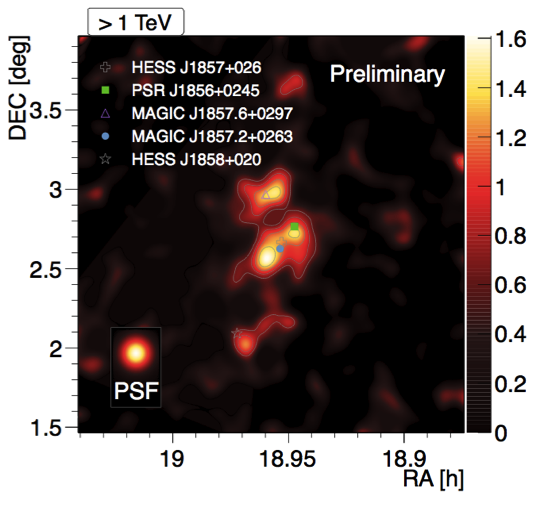}
  \caption{MAGIC gamma-ray flux map of 
    HESS~J1857+026 for events with estimated energy greater than 1
    TeV, resolving the extended source into two distinct components
    \cite{MAGIC_HESSJ1857}.}
  \label{MAGIC_HESSJ1857_fig}
\end{figure}

Given their prevalence among TeV sources in the Galactic plane,
coupled with emission spatially offset from the host pulsar location,
PWN studies at TeV energies often involve a certain amount of
detective work. Two particularly interesting ``cases'' were solved in
presentations at this ICRC. MAGIC observations of HESS J1857+026
resolved this extended source into two components at energies above
$1\U{TeV}$. One of these is likely the elongated PWN associated with the young
energetic pulsar PSR~J1856+0245. The other appears point-like, and remains unidentified
\cite{MAGIC_HESSJ1857} (Figure~\ref{MAGIC_HESSJ1857_fig}. 
VERITAS presented the results of a deep ($\sim50\U{hour}$) observation
of TeV~J2032+4130 \cite{VERITAS_Cygnus}, the first unidentified source
discovered at very high energies (by HEGRA \cite{HEGRA_TeV2032}). The
source is determined to be extended, with an asymmetric morphology
which, in the context of multiwavelength observations of the same
region, favours a PWN interpretation, associated with the LAT pulsar
PSR~J2032+4127.

\begin{figure}[!]
  \centering
  \includegraphics[width=0.48\textwidth, clip,trim = 0mm 0mm 1mm 0mm]{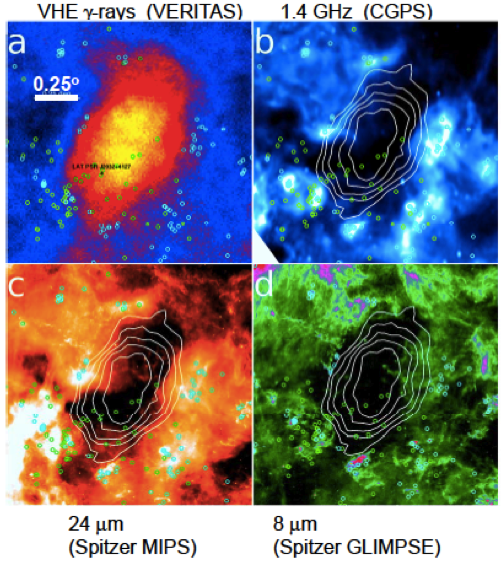}
  \caption{The VERITAS skymap for TeV~J2032+4130, along with
   radio and infra-red observations of the same region \cite{VERITAS_Cygnus}.}
  \label{VTS_TeV2032_fig}
\end{figure}

\subsubsection*{Supernova Remnants}

The most exciting result in gamma-ray supernova remnant (SNR) studies
since the last ICRC is undoubtably the discovery of direct evidence
for the existence of hadronic cosmic ray particles in W~44 and IC~443,
through the characteristic pion-decay signature in their sub-GeV
spectra with Fermi-LAT \cite{Funk_pion}. However, the acceleration and
escape processes are not yet well understood, and it is still far from
clear whether particle acceleration in SNRs can explain the bulk of
the Galactic cosmic ray population up to energies beyond the
knee. Ground-based observations could help to resolve these questions,
and supernova remnant results presented at the conference showed
progress through some new discoveries and  deeper studies of
established sources, as well as initial results from an unbiased
population study \cite{HESS_SNR_pop}.

\begin{figure}[!]
  \centering
  \includegraphics[width=0.48\textwidth]{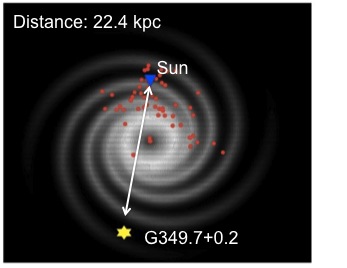}
  \caption{A cartoon showing the location in the Galactic diskof all known Galactic TeV
    sources (red), plus the newly discovered HESS source
    SNR~G349.7+0.2 \cite{HESS_distant_SNR}.}
  \label{HESS_SNR_G349}
\end{figure}

H.E.S.S. presented observations of a new TeV source associated with
the bright young SNR~G349.7+0.2. The source received a deep exposure
($\sim100\U{hours}$) as it shares the same field-of-view as
RX~J1713.7-3946, allowing a detection despite its location on the far
side of the Galaxy (at a distance of $\sim22.4\U{kpc}$)
\cite{HESS_distant_SNR} (Figure~\ref{HESS_SNR_G349}). The emission is
satisfactorily explained within a hadronic
interpretation. HE.S.S. also presented a reinterpretation of the most
luminous Galactic TeV source, HESS~J1640-465, which was previously
identified as a likely PWN. The absence of an inverse Compton peak in
the broadband gamma-ray spectrum, combined with a TeV morphology which
overlaps with the northern rim of the SNR, now appears to favour
hadronic emission from the SNR shell \cite{HESS_bright_SNR}.

Resulting from a historical Type Ia progenitor supernova explosion,
Tycho's SNR is a shell-type remnant expanding into the undisturbed
interstellar medium. The relative simplicity of this system has made
it a favorite choice for modellers, and the detection of TeV and GeV
emission was shown at the last ICRC. In Rio, updated results from
VERITAS extended the lower end of the spectrum to $300\U{GeV}$
\cite{VERITAS_Tycho}. The statistical errors still accommodate a
variety of different models (leptonic and hadronic, one- and
two-zone), but data collection is ongoing. The W51 region, on the
other hand, is a complex environment including a supernova remnant
(W51C) interacting with the molecular clouds of the star-forming
region W51B, as well as a PWN candidate. A TeV source in this region
was originally detected by H.E.S.S. At this meeting, MAGIC presented
the results of a $50\U{hour}$ exposure which extends the spectrum, and
appears to indicate that the majority of the emission in the region
comes from the SNR-molecular cloud interaction \cite{MAGIC_W51}.

Finally, an interesting (if speculative) search for gamma-ray emission
associated with \textit{very} young nearby \textit{extragalactic} SNR
was presented \cite{HESS_recent_SN}. Such objects might possibly
provide the source of particles to between the knee and the angle of
the cosmic ray spectrum. A search for serendipitous
H.E.S.S. observations of nearby ($z<0.01$) supernova locations within
$1\U{year}$ of the explosion yielded 9 candidates, but no detections.

\subsubsection*{Gamma-ray Binary Systems}

The binary source class remains relatively small, at both GeV and TeV
energies, but displays a rich phenomenology. The lightcurve of the
only binary to be first discovered from the ground, HESS~J0632+057 has
now been measured over almost a decade's worth of 315-day orbits by
VERITAS and H.E.S.S., showing close correlation between the TeV and
X-ray flux, and including the first detection of TeV emission at
phases outside of the bright X-ray flare phase \cite{HESS_binaries,
  VERITAS_binaries}. The measurement of variability in
HESS~J1018–589~A now firmly associates this point-like component of an
extended TeV source with 1FGL~J1018.6-5856, the $16.6\U{day}$ binary
system discovered by Fermi-LAT \cite{HESS_binaries}.

\begin{figure}[!]
  \centering
  \includegraphics[width=0.48\textwidth, clip,trim = 0mm 0mm 1mm 0mm]{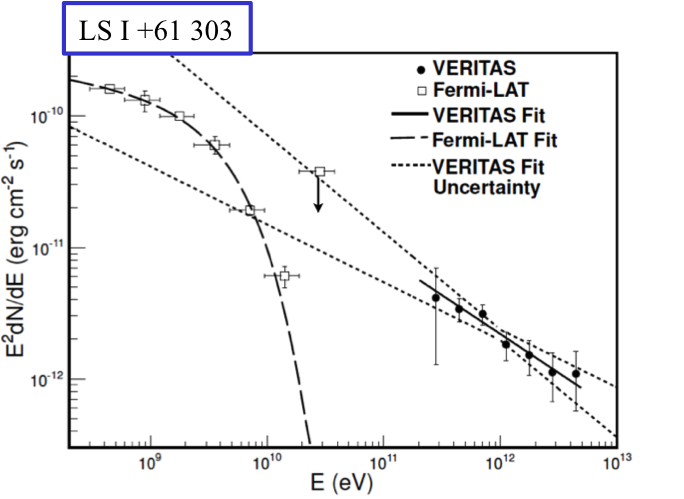}
  \caption{The spectral energy distribution of LS~I+61$^{\circ}$303
    from contemporaneous Fermi-LAT and VERITAS observations in
    2011/2012, showing a distinct gap in the emission between ∼4 GeV
    and 300 GeV \cite{VERITAS_binaries}.}
  \label{LSI_fig}
\end{figure}

LS~I+61$^{\circ}$303 is among the most heavily observed TeV sources in
the northern hemisphere, but the results continue to surprise. New
observations from VERITAS and MAGIC since the last ICRC show that the
source has been bright at phases close to apastron for the first time
since the launch of Fermi-LAT \cite{VERITAS_binaries,
  MAGIC_binaries}. This allows to construct a contemporaneous
high-energy spectrum, revealing a sharp cut-off between the GeV and
TeV spectra (Figure~\ref{LSI_fig}). The explanation for this ``gap'' in
the spectral energy distribution (SED) is unclear, but it seems to indicate
that the emission mechanism may be different in the GeV and TeV
bands. The re-emergence of bright emission from LS~I+61$^{\circ}$303
also led to speculation that the high-energy emission may be modulated
with the $\sim5\U{year}$ super-orbital period observed in X-ray and
radio, which may be related to changes in the circumstellar disk of
the Be star.

\subsubsection*{Globular Clusters}

The detection of GeV emission from numerous globular clusters by
Fermi-LAT, and of TeV emission from the direction of Terzan 5 by
H.E.S.S. \cite{HESS_T5}, motivates the search for other members of this source
class. Emission might originate from the combined emission of the
numerous millisecond pulsars in the cluster cores, or from inverse
Compton scattering of relativistic leptons accelerated in the
cluster environment, for example close to the aforementioned
millisecond pulsars. The Terzan 5 association is not definitive,
primarily because the TeV source is slightly offset from
the cluster center. The addition of another TeV/ cluster association
would make the identification much more compelling. Results of a
search for TeV emssion from 15 globular clusters were presented by
H.E.S.S. \cite{HESS_glob}, which revealed no new detections.

As a final point on Galactic sources, it is worth noting that one of
the largest populations of TeV gamma-ray sources remains unidentified
objects in the Galactic plane.  For many of these, the search for a
definitive counterpart is still ongoing, requiring ever deeper
exposures and multiwavelength detective work (e.g.\cite{HESS_J1832,
  VERITAS_Cygnus}).

\section*{Results: Extragalactic Sources}
\subsubsection*{Blazars}  
\begin{figure*}[!t]
  \centering
  \includegraphics[width=\textwidth]{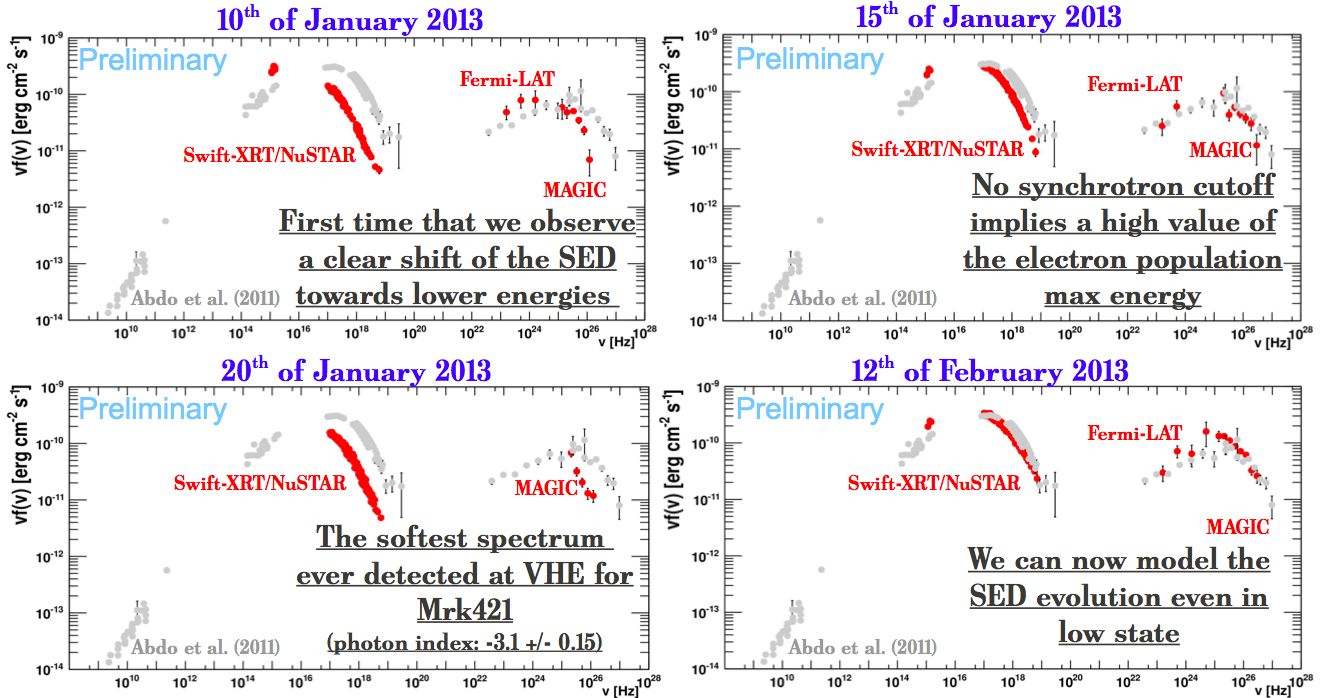}
  \caption{Daily measurements of Markarian 421 in different states in
    2013, including the first contemporaneous observations with
    NuSTAR, which observes in the 3-$79\U{keV}$ energy range
    \cite{421_2013_flare}. The grey points show an average spectrum
    from \cite{421_average}}
  \label{Nustar_421}
\end{figure*}

The TeV blazar population continues to grow and, with the addition of
five new members to the class announced in the run-up to this ICRC
(KUV~00311-1938, PKS~0301-243, PKS~1440-389 \cite{HESS_status};
H1722+119, MS~1221.8+2452 \cite{MAGIC_status}), now numbers 49
sources. One development at the ICRC was the first presentation of
broad-band blazar spectral energy distributions including data from
NuSTAR, a focussing X-ray telescope launched in June 2012, which
operates in the 3-$79\U{keV}$ energy range
\cite{421_2013_flare}. These observations, combined in coordinated
campaigns with observations at lower and higher energies, now allow to
characterize in great detail both the synchrotron and inverse Compton
emission components of nearby blazars on a \textit{daily} basis, even
when the source is relatively quiescent. Prior planning is generally
required for these campaigns, and so the fortuitous detection of a
bright flare (exceeding 11 Crab in the TeV band) from Markarian 421 in
April 2013 during multiwavelength observations is very exciting. Only
preliminary results for an earlier part of the campaign were shown at
the ICRC (Figure~\ref{Nustar_421}), and the interpretation is still in
the earliest stages, but even these provide useful new information,
including the absence of any evidence for a cut-off in both components
of the SED.

More complete modelling has been applied to observations of Markarian
421 from an earlier, 2010, multiwavelength campaign, which captured a
decaying flare from the source \cite{421_2010_flare}. The authors
model this with both one and two-zone synchrotron self-Compton
emission models, and conclude that the evolution of the SEDs favours
the presence of two 'blobs' of material, rather than the single blob
typically used to describe flares in TeV blazars. A summary of
H.E.S.S. observations of blazars similarly concluded that simple
one-zone SSC models rarely provide a satisfactory fit to the complete
SED, and are also unable to explain the multi-wavelength variability
patterns \cite{HESS_AGN}. We now appear to be at a healthy stage in
blazar studies, where the quality of the available observations
exceeds theoretical interpretations. The data quality will further
improve in the coming years as continuous monitoring with survey
instruments such as HAWC (and, to a lesser extent, dedicated
monitoring with FACT) will help to catch high energy flares and to
better characterize the temporal variability.

Gamma-ray emission from blazars can also be used as a probe of the
intensity of the extra-galactic background light (EBL), due to the
absorption of gamma-ray photons by pair-production. My co-rapporteur
summarized published work by H.E.S.S. in which a clear detection
($8.8\sigma$) of the imprint of the EBL was detected in a stacked
analysis of 75,000 gamma-ray photons from 7 bright blazars with
$z<0.19$ \cite{Abramowski_EBL}. The most exciting new results in this
area at the ICRC concerned two sources at the gamma-ray horizon.
Hubble Space Telescope observations of hydrogen line absorption
features in the optical spectrum of PKS~1424+240 have been used to
place a firm lower limit on its redshift of $>0.6035$, making this the
most distant TeV source to be detected \cite{Furniss_1424}. The
results of new VERITAS observations of this source were shown
\cite{VERITAS_AGN}, as were MAGIC observations of another very distant
TeV blazar, PG~1553+113, at $z>0.4$ \cite{MAGIC_1553}.  For
PG~1553+133, spectral curvature is observed, which can be explained as
due to the energy dependent absorption by the EBL. For both
PG~1553+133 and PKS~1424+240, the intrinsic spectra, corrected for
absorption effects, appear unusually hard, with some indication of an
upturn at the high-energy end of the spectrum.  The strength of this
feature depends on the EBL model which is applied, however, and stronger
conclusions could be drawn with both better statistics at the highest
energies and a firm redshift determination for each source.

Flat-spectrum radio quasars (FSRQs) are almost as common as BL Lac
objects in the Fermi catalog, but comprise only a small subset of
extragalactic TeV sources (3 objects), due to their soft, steep
spectra and generally larger distances. MAGIC observations of
PKS~1510-089 resulted in a confirming detection of this source, first
detected by H.E.S.S \cite{HESS_AGN, MAGIC_FSRQs}, while 3C~279, the first,
and most distant ($z=0.5362$) FSRQ to be detected at TeV energies, has
also been re-observed, but not detected. The origin of the TeV
gamma-ray flares, and the correlation between the TeV emission and
emission at other wavelengths, remains poorly defined for these sources.

\subsubsection*{Radio Galaxies}

Observations of nearby radio galaxies, in which the jet is not
directly oriented towards the line-of-sight, are also of high priority
for the TeV observatories. In particular, these allow attempts to
correlate TeV emission properties with changes in the internal
structure of the jet, which can be resolved at other wavelengths.

\begin{figure}[!]
  \centering
  \includegraphics[width=0.48\textwidth, clip,trim = 0mm 0mm 1mm 0mm]{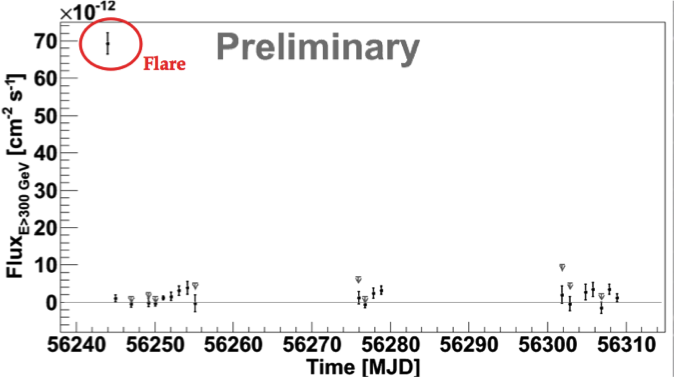}
  \caption{The TeV lightcurve of IC~310, showing a bright flare ($>0.5$ Crab) detected
    in November 2012 \cite{MAGIC_IC310}.}
  \label{IC310}
\end{figure}

IC~310 was originally classified as a head-tail radio galaxy, a class
of objects characterized by their extended jets which point in a
direction determined by the galaxy's motion through an intra-cluster
medium. High energy emission was detected from this object by Fermi
above $30\U{GeV}$, and during MAGIC observations in 2009/10, as
reported at the time of the last ICRC \cite{Neronov_IC310,
  MAGIC_IC310_ApJ}. Since then, a one-sided blazar-like compact radio jet has
been identified in VLBI radio observations, which argues against the
head-tail classification, and suggests that IC~310 may instead be the
closest known blazar ($z=0.0189$). MAGIC observations in November 2012
detected an exceptionally bright flare, lasting less than one day and
reaching a level in excess of half of the Crab Nebula flux above
$1\U{TeV}$ \cite{MAGIC_IC310} (Figure~\ref{IC310}). The TeV spectrum
during the flare is unchanged, and radio monitoring does not yet, at
the time of this meeting, show any response of the radio jet to the
gamma-ray flare. Monitoring continues, at radio and other wavelengths,
since the typical timescale between a gamma-ray flare and the ejection
of new radio components may be several months or longer.

Separated from IC~310 by just $0.6^{\circ}$ on the sky is the central dominant
galaxy of the Perseus cluster, NGC~1275. With a visible counter-jet,
NGC~1275 is one of only three, non-blazar AGN detected above
$100\U{GeV}$ (the others being M~87 and Cen~A)
\cite{MAGIC_NGC1275_AA}. The power-law source spectrum breaks sharply
at GeV energies, becoming one of the softest spectrum sources to be
detected from the ground at higher energies ($\Gamma \sim -4.0$). A
confirming detection of the source with VERITAS was only possible
following a substantial reduction in the energy threshold of the
upgraded instrument \cite{VERITAS_status}. Long term monitoring at
TeV, GeV, optical and radio wavelengths was also presented at this
conference and shows correlated emission between the optical flux and
that above $>100\U{MeV}$ \cite{MAGIC_NGC1275}, suggesting a common
point of origin in the core of the AGN. The broadband spectrum is
well-described by a simple one-zone synchrotron self-Compton model, as
might be expected for a BL Lac type object with a misaligned jet.

By far the most well-studied of the TeV radio galaxies is M~87, which
is easily visible from both the Northern and Southern hemispheres. No
new results were shown at this meeting, but a summary of 10
years of observations with gamma-ray and multiwavelength partners
illustrates that, despite the exceptionally rich dataset, a unique
signature linking TeV emission with flares or jet structures at other
wavelengths is still yet to be established \cite{M87, M87_ApJ}. 

\section*{Other results in Astroparticle Physics}

Ground-based gamma-ray observations have been used in a variety of
sometimes innovative ways to address other questions in astroparticle
physics, fundamental physics and cosmology. At this meeting, a search
for modifications to the intrinsic gamma-ray spectrum of the bright
blazar PKS2155-304 was used to constrain the strength of the coupling
of axion-like particles to photons \cite{HESS_Axions}. An analysis of
the full H.E.S.S. data archive was used to improve the upper limit on
the local rate of primordial black hole explosions by almost an order
of magnitude, to $<1.4\times10^{4}\UU{pc}{-3}\UU{yr}{-1}$
\cite{HESS_PBH}. Searches for evidence of Lorentz invariance violation
have customarily focussed on transient events, such as blazar flares
or gamma-ray bursts, but pulsars also provide a high-energy photon
flux with a sharp temporal profile. The VERITAS collaboration have
exploited their observations of the Crab pulsar to constrain the
energy scale of Lorentz invariance violations to
$>3\times10^{17}\U{GeV}$ \cite{Zitzer_LIV}. The sensitivity of the
full HAWC observatory to address questions of this nature has also
been studied, with promising conclusions \cite{HAWC_LIV, HAWC_PBH}.
 
The changing scientific priorities in astroparticle physics warranted
the creation of a dedicated dark matter (DM) sessions for this
meeting, distinct from the gamma-ray sessions. DM submissions,
including indirect detection techniques, were comprehensively reviewed
by the DM session rapporteur. Results related to ground-based
gamma-ray instruments include new limits on WIMP annihilation
cross-sections from observations of dwarf spheroidal galaxies
\cite{Zitzer_DM, MAGIC_Segue, HESS_Sag_dwarf}, along with the
description of new analysis techniques for such datasets
\cite{Rico_DM}. Methods for ``stacking'' the results from observations
of multiple source candidates holds promise, and may alleviate the
potential for systematic errors associated with extremely long
exposures on a single field-of-view. The development of statistical
methods to achieve this can also spread the observing load over
multiple instruments, and allow to factor in complementary information
from adjacent wavebands. Observations of a Fermi-LAT dark matter
subhalo candidate were also described \cite{MAGIC_UID}, as was a
search for the signature of decaying DM \cite{Decaying_DM} and the
potential for detecting Q-balls with HAWC \cite{HAWC_Qballs}.

\section*{The Future}

Major upgrades to all of the current generation of IACTs have recently
been completed, and new facilities are now in the prototyping and
advanced planning stages. This growth in development activity was
reflected at the ICRC in a large number of detailed technical
presentations, many of which were presented in the poster sessions.

\begin{figure}[!]
  \centering
  \includegraphics[width=0.48\textwidth]{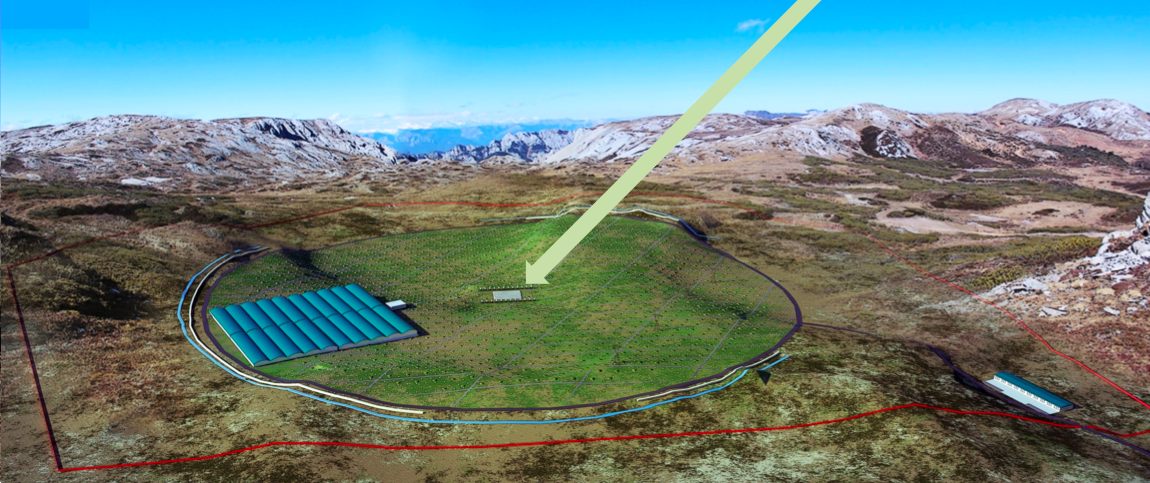}
  \caption{A schematic view of the proposed LHAASO array. See text and
  Cao et al. \cite{LHAASO} for details.}
  \label{LHAASO_layout}
\end{figure}

The Large High Altitude Air Shower Observatory (LHAASO) collaboration
are proposing an ambitious survey instrument to study both cosmic rays
and gamma-rays, which builds on the success of the ARGO-YBJ and Tibet
arrays (Fig.~\ref{LHAASO_layout}) \cite{LHAASO}. The first phase of this project will consist of:

\begin{itemize}
\item{A $1\UU{km}{2}$ array of 5635, $1\UU{m}{2}$  scintillator
    detectors, with $15\U{m}$ spacing, for electromagnetic particle detection \cite{LHAASO_KM2A}}
\item{An overlapping $1\UU{km}{2}$ array of 1221, $36\UU{m}{2}$ underground water
    Cherenkov tanks, with $30\U{m}$ spacing, for muon detection \cite{LHAASO_KM2A}}
\item{A close-packed, surface water Cherenkov detector facility with a total
    area of $90,000\UU{m}{2}$, four times that of HAWC}
\item{24 wide field-of-view air flourescence (and Cherenkov) telescopes}
\item{425 close-packed burst detectors, for high energy secondary particles near
    the centre of the array}
\end{itemize}

Various prototyping studies of the different detector components and
associated instrumentation were presented at the ICRC, including the
successful demonstration of a small engineering array at Yangbajing.
A site for the full array has been selected at Shika, close to
Yangbajing, and at a similar altitude of 4300m above sea level. The
project still has some remaining administrative hurdles to cross but,
once funding is secured, construction will begin (in 2014), and last
approximately 5 or 6 years. Also in the realm of the survey
instruments, possible ``third generation'' successors to HAWC were
discussed, with potential improvements including higher altitude,
larger active collection area, separate measurement of electromagnetic
and muon components of the shower, and a southern hemisphere site to
improve access to the Galactic Center region \cite{HAWC_NG}.

The most advanced development activities towards a next generation
ground-based gamma-ray telescope were shown by the Cherenkov Telescope
Array (CTA) collaboration, with around 60 presentations at this
ICRC. Almost the entire international ground-based community has now
united behind this effort, and the project has moved on significantly
since the last ICRC. I will briefly list some of the major advances;
Jim Hinton's review talk at this conference on the status and future
of gamma-ray astronomy also contains more information on CTA \cite{Hinton_review}.

CTA is planned to consist of a large array in the southern hemisphere
and a smaller array in the north. The sites have been evaluated for
their suitability for Cherenkov observations using a variety of
different sky and environmental monitoring tools
\cite{CTA_site_survey, CTA_atmoscopes, CTA_ASCs}. At the time of
writing, formal negotiations on two sites in the south (in Namibia and
Chile) are underway, with an Argentinian site \cite{CTA_Argentina} in
reserve as a third option. No decision has been made regarding the
northern site, but various candidates have been proposed at locations
in Spain \cite{CTA_Spain}, Mexico and the USA \cite{CTA_USA}.

\begin{figure*}[!]
  \centering
  \includegraphics[width=0.8\textwidth]{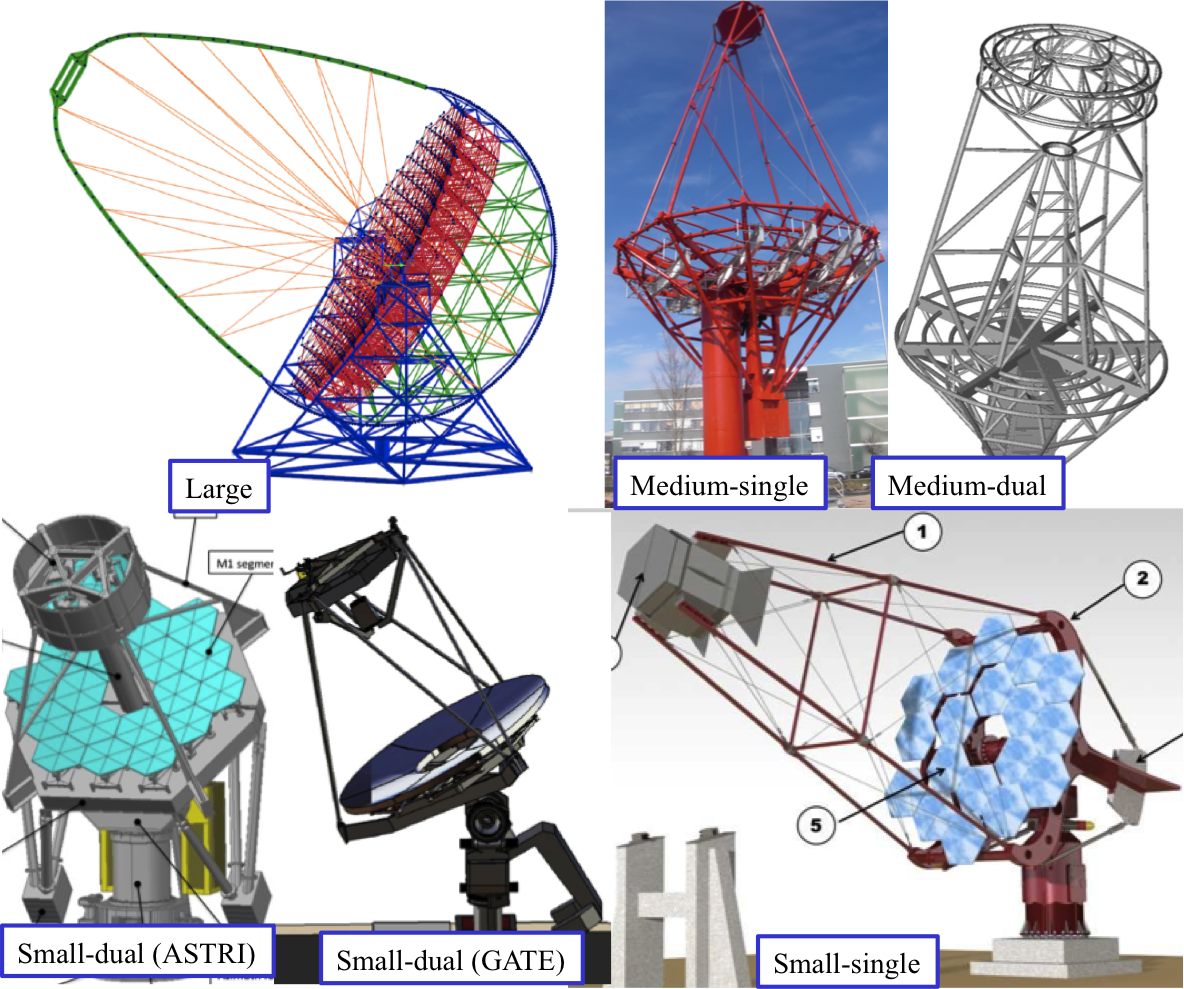}
  \caption{A summary of the different telescope designs in
    development for CTA.}
  \label{CTA_tels}
\end{figure*}

The arrays themselves will be comprised of telescopes of multiple
different designs, to optimize the sensitivity and to provide the
widest possible coverage in energy (Figure~\ref{CTA_tels}). At the
center of the array will be four ``Large Size Telescopes'' (LSTs)
\cite{CTA_LST}, with reflector dish diameters of $23\U{m}$, optimized
for $\sim20-200\U{GeV}$. The design for these telescopes is based upon
the succesful MAGIC design, scaled up in size, but down in weight and
cost, through the application of new technologies and materials. The
camera will consist of 2000 super-bialkali PMTs, read-out using the
DRS4 analog memory ASIC \cite{CTA_LST_Camera, CTA_LST_Readout}.

``Medium Size Telescopes'' (MSTs), will provide deepest coverage of
the mid-range of energies, centered around $1\U{TeV}$. Two designs are
envisaged for the MSTs, both of which may be deployed. The first is a
single reflector Davies-Cotton design, similar to the VERITAS and
original H.E.S.S. telescopes, with a diameter of $12\U{m}$. The second
is a novel dual-mirror design, using a Schwarzchild-Couder optical
system, with a $9.5\U{m}$ primary reflector. The single reflector MST
exists in an advanced prototype form at the Adlershof test site near
Berlin. The dual mirror version is under development and will be
prototyped at the VERITAS site in Arizona \cite{CTA_SCT}.

The high energy range of the array, up to hundreds of TeV, will be
covered by the ``Small Size Telescopes'' (SSTs). These telescopes,
with an aperture of $\sim4\U{m}$, will be the most numerous, and most
widely spaced. Two dual-mirror designs \cite{CTA_SST_2M, CTA_SST_GATE,
  CTA_SST_ASTRI}, and one single mirror system \cite{CTA_SST_1M} are
in the prototyping stage, and will be tested at sites in Poland, Paris
and Italy.

\begin{figure*}[!]
  \centering
  \includegraphics[width=0.7\textwidth]{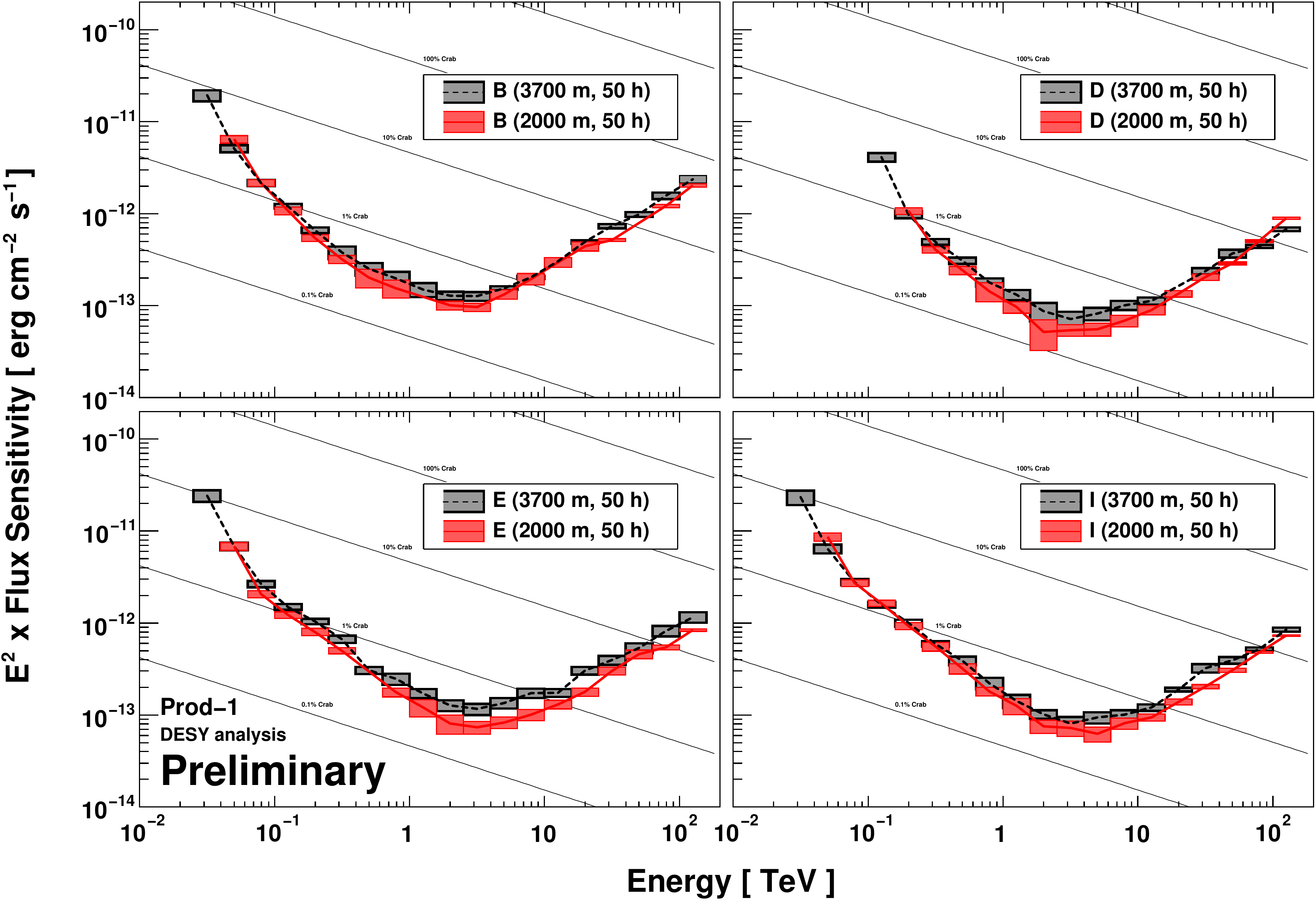}
  \caption{The differential sensitivity of for four candidate CTA array
    configurations at two different altitudes (red: $2000\U{m}$,
    black: $3700\U{m}$. Figure from \cite{CTA_sims}.}
  \label{CTA_sens}
\end{figure*}

In parallel with the telescope structural design development, advanced
prototyping and testing is underway in mirror designs
\cite{CTA_mirrors}, and telescope camera, trigger and data acquisition
systems (e.g. \cite{CTA_LST_Readout, CTA_CHEC, CTA_FlashCam,
  CTA_dacq}). New technologies are being tested at all stages, both in
order to enhance the array performance, and to deal with the
necessities of mass production, low cost, and strict maintenance
requirements.  (e.g. dielectric mirror coatings, high quantum
efficiency and multi-anode PMTs, silicon-based photodetectors, etc.).
A significant effort is also being put into peripheral and monitoring
systems and calibration efforts (e.g. \cite{CTA_atmo, CTA_pointing,
  CTA_laser}). Monte Carlo simulations \cite{CTA_sims} and
data analysis tools are also well advanced, and are being used to
inform design decisions, including the possible array
layouts and the impact of different observatory locations (Figure~\ref{CTA_sens}).

\section*{Concluding remarks}

Every ICRC over the last decade has seen dramatic advances in
ground-based gamma-ray astronomy, and this meeting was no
exception. Although the high energy sky continues to surprise us, it
would be fair to say that the most significant advances shown in Rio
were in instrumentation, other than observational or
theoretical. Upgrades to existing facilities, new facilities coming
online, and intensive development work towards the next generation of
instruments were all presented. The next ICRC will showcase the first
fruit of these efforts and promises to mark another important step in
the development of the field. I look forward to seeing you all there.

\section*{Acknowledgements}

I would like to thank the organisers of the ICRC and, in particular,
the local support in Brazil, for hosting such a well-organised and
enjoyable meeting. The ICRC 2013 is funded by FAPERJ, CNPq, FAPESP,
CAPES and IUPAP.


\end{document}